# Influence of Dynamics on Magic Numbers for Silicon Clusters

A. R. Porter[*] and R. W. Godby[†]

Department of Physics, University of York, York YO1 5DD, United Kingdom

**ABSTRACT:** We present the results of over 90 tight-binding molecular-dynamics simulations of collisions between three- and five-atom silicon clusters, at a system temperature of 2000K. Much the most likely products are found to be two 'magic' four-atom clusters. We show that previous studies, which focused on the equilibrium binding energies of clusters of different sizes, are of limited relevance, and introduce a new effective binding energy which incorporates the highly anharmonic dynamics of the clusters. The inclusion of dynamics enhances the magic nature of both $Si_4$ and $Si_6$ and destroys that of $Si_7$.

In recent years much work has been done on developing the understanding of the structure of atomic clusters and their transition to the structure of a bulk solid with increasing size. For silicon in particular, the existence of magic number clusters, consisting of 4, 6 and 10 atoms, has been the subject of much study using both *ab initio* and tight-binding (TB) methods [1,2,3]. These magic numbers correspond to especially abundant cluster sizes arising during the production process, where laser evaporation of a solid silicon surface produces a plasma which is then swept away in a stream of inert gas. As this plasma cools, the clusters undergo a series of collisions as the system attains dynamic equilibrium. It is found [4] that the magic numbers indicated by the mass spectrum of the clusters so produced are independent of the details of the method of production, depending only on the fact that the clusters have reached dynamic equilibrium.

Previous work [1,2,3,5] has attempted to explain the existence of these magic numbers through the study of the equilibrium structures of the clusters. However, during their formation the clusters are at a high temperature and are consequently exploring geometrical configurations far from the potential-energy minimum. The relevance of the minimum energy is therefore unclear. The aim of this work is to simulate a large number of collisions, analyzing the results in terms of the *dynamics* of the reactants, intermediate clusters and products, rather than simply the statics of cluster equilibrium.

Use of a tight-binding (TB) scheme makes such simulations computationally feasible whilst incorporating the quantum-mechanical description of the *sp* directional bonding in silicon that is essential for reproducing the correct cluster structures. The TB scheme used here was devised by Goodwin *et al.* [6] to give a good representation of a large database of the volume-dependent total energy of solid silicon in various structures, ensuring accuracy over a wide range of bonding configurations. The energy and forces obtained from the TB model were used in a classical molecular-dynamics (MD) simulation [7] of the atomic motion.

Since all previous work has concentrated on the equilibrium structures of the Si clusters, we address these first. Hence, we used our TB MD, to perform simulated annealing for clusters of 2 to 8 atoms in order to identify their minimum-energy structures. All of the structures identified in this way were found to agree with previous results using this scheme [6] except for the case of $Si_5$. For this cluster we find a rather asymmetric, face-capped planar rhombus form for the minimum energy structure, in agreement with the results of Laasonen and Nieminen [5]. It is significant that the search for the equilibrium configuration of $Si_8$ was quite difficult, it being found necessary to reduce the time step to one twentieth of the $2.42 \times 10^{-4}$ ps used during the cooling of all of the other clusters. This indicates that the $Si_8$ minimum energy structure is a narrow and rather inaccessible valley in the potential energy surface of the cluster.

Although the TB scheme used here achieves better agreement with *ab initio* results for the equilibrium structures of the clusters than other TB schemes [1,2] (typically bond lengths are within 10% of *ab initio* values), the associated binding energies per atom were consistently overestimated by $(1.03 \pm 0.07)$ eV. This may be due in part to the fact that Goodwin *et al.* [6] fitted the scheme to the results of an *ab initio* density-functional-theory calculation in the local-density approximation (LDA), since calculations of this type overestimate the cohesive energies of bulk silicon and molecules. However, as the overbinding is virtually independent of cluster size (and thus of the different types of bonding present in the different structures), it is unlikely to have had a significant effect on the simulated dynamics and the outcomes of the collisions. It should be noted that these errors are of the same order of magnitude as would be obtained from an *ab initio* LDA calculation.

The results of the simulated annealing indicated that the equilibrium structure of the $Si_4$ cluster was considerably more stable than either the 3 or 5 atom forms. By simulating the collision of $Si_3$ with $Si_5$, we aim to investigate the effects, if any, of this property of $Si_4$ on the outcome of the cluster-cluster collision process. The collisions were carried out for impact parameters, *b*, of 0 to 5.82 Å with an initial system temperature of approximately 2000K and were followed for a time corresponding to 7.26 ps. Although locating the minimum energy structure of $Si_8$ required a much reduced time step, its highly localized nature means that it will not significantly affect the dynamics during the collisions simulated here [8]. It is therefore acceptable to use the standard time step of $4.84 \times 10^{-4}$ ps. Prior to a collision, the clusters were separately heated to 2000K and assigned appropriate centre-of-mass velocities. In order that the results may be considered as an average over the many possible relative orientations of the colliding clusters, we rotated each cluster through three random angles about its internal axes before every collision.

A summary of the results of all of the collisions simulated is given in Table I. For all cases where the clusters came into contact a reaction resulted, such as the transfer of a single atom to produce $Si_4 + Si_4$. Although twelve simulations were performed for each impact parameter, the clusters failed to make contact on three occasions in simulations with the largest value of $b$ and thus did not react. The eight reactions that produced $Si_3 + Si_5$ all involved significant rebonding.

The results in Table I clearly show the dominance of the $Si_4 + Si_4$ reaction product in these collisions, with this configuration being the result in almost 70% of the reactive collisions. 20% of the reactions produced a single $Si_8$ cluster that did not fragment, making this the next most abundant product. It is interesting to note that the loss of a single atom by the highly mobile $Si_8$ cluster formed during the collisions was never observed.

We emphasize the irrelevance of the *equilibrium* potential energy values of the clusters in determining the outcomes of these collision reactions. As an alternative, we have calculated the *effective* binding energies (EBEs) of the clusters under the conditions simulated here. The effective binding energy was defined to be −1 times the minimum value of potential energy that would exist *if* the potential-energy surface for the *actual* mean potential energy and kinetic energy were harmonic about its minimum (see Figure 1). Since, for a harmonic oscillator, $\langle PE_{vib} \rangle = \langle KE_{vib} \rangle - BE_{eff}$, (where $\langle \ \rangle$ indicates a time average) and $\langle KE_{vib} \rangle$ can be calculated from the total kinetic energy by using the equipartition theorem, this assumption enables an estimate of the minimum binding energy to be obtained from the simulation [9]:

$$BE_{eff} = -\left[ \langle PE_{vib} \rangle_{sim} - \frac{(3N-5)}{3N} \langle KE \rangle_{sim} \right] \quad (1)$$

where $N$ is the number of atoms in the cluster. All of the time averaged values except that for $Si_7$ were obtained from the results of actual collision runs. Since $Si_7$ was never produced in these collisions, its effective binding energy was obtained by heating an example to approximately 4000K (typical of the temperature attained by $Si_8$ during a reaction) and then letting it evolve in isolation. Since a cluster's temperature determines the extent to which it explores what is, in reality, an anharmonic potential energy surface, the appropriate EBE to characterize its motion will also be a function of temperature. Therefore, because the temperature of the reacting $Si_3$ and $Si_5$ clusters tended to differ from the temperatures of such clusters produced during a reaction, two values for the EBEs of these clusters have been calculated.

The resulting EBEs differed significantly from the corresponding equilibrium values in a number of ways. For instance, the EBEs of both $Si_7$ and $Si_8$ are considerably less than the equilibrium values, demonstrating the inaccessibility of their equilibrium structures in the dynamics simulated here. In contrast, for $Si_4$, the effective binding energy is greater than the equilibrium binding energy, indicating a flattening, rather than a steepening, of the potential-energy surface near its minimum.

The energy changes associated with different reaction outcomes may be estimated by subtracting the binding energies of the reacting $Si_3$ and $Si_5$ clusters from those of the products. The estimates obtained with both the equilibrium and effective values for the cluster binding energies predict $Si_8$ to be the most favourable product, followed by $Si_4 + Si_4$. However, the difference between these two products is much less in the EBE estimate (1.3 eV compared with the equilibrium value of 3.9 eV). In general, the estimates produced using the EBE's give predictions in much closer agreement with both the results of our simulations and those of experiment [4] than do the corresponding equilibrium estimates. First, use of the effective values reduces the predicted increase in binding energy associated with the production of $Si + Si_7$ from 1.7 eV (the equilibrium estimate) to −0.9 eV. This indicates that it is not an energetically-favourable outcome; hence its absence from Table I. Second, they also give the correct ordering of the remaining products in that $Si_3 + Si_5$ is predicted to be marginally (0.1 eV) more favourable than $Si_2 + Si_6$, in direct contrast to the equilibrium values which predict that $Si_2 + Si_6$ is the more favourable of the two by 0.7 eV. The fact that a product of $Si_8$ is not as common as suggested by this simple estimate is due to the fact that the single cluster cannot accommodate all of the kinetic energy of the reaction. Hence, it becomes simply a matter of time as the cluster explores its configuration space before it will fragment.

Another way of comparing the equilibrium and effective binding energies is to plot them per atom as a function of cluster size (Fig. 2). The additional stability of the points at $N=4$ and $N=6$, in comparison to what would be expected from a linear combination of neighbouring values, indicates the magic status of $Si_4$ and $Si_6$. In contrast, the corresponding curve for the equilibrium values has no local maximum at $N=6$, predicting instead that $Si_7$ is magic. These features are emphasized in the plot of cluster fragmentation energy (defined as the binding energy of an $N$-atom cluster less that of the $N-1$-atom cluster) also shown in Figure 2. The peaks at $N=4$ and $N=6$ in the curve obtained with the effective values show the enhanced stability of these clusters under the dynamic conditions simulated here. The absence of a peak for the $Si_7$ cluster reflects the removal of its magic nature.

As a typical example of the type of the behaviour observed during the collision processes we shall briefly discuss a particular collision with an impact parameter of 5.82 Å. The variation with time of the system potential energy and mean coordination number during this collision is shown in Figure 3. (The definition of coordination for the latter quantity included a linear cut-off, chosen to maintain a nearest-neighbour character and thus a coordination number of 4 in diamond-structure bulk silicon.) The time of first contact of the clusters is indicated by an arrow in the figure and corresponds to the sharp drop in PE as the clusters begin to mix. It should be noted however, that the new $Si_8$ cluster never attains its minimum energy of $-37.2$ eV during the mixing phase, again emphasising the fact that its equilibrium structure is not significant under these conditions. At $t=3.8$ ps, the mean coordination number drops briefly to its initial value as the $Si_8$ almost fragments into $Si_3 + Si_5$. However, the clusters recombine before finally splitting (indicated by second arrow) as $Si_4 + Si_4$ at approximately 4.8 ps. The final potential energy of the system is considerably lower (by roughly 1.4 eV) than that of the reactants.

In conclusion, we have simulated over ninety collisions of $Si_3$ with $Si_5$ at a system temperature of approximately 2000K. Over 70% of the reactive collisions resulted in the production of $Si_4 + Si_4$, whereas $Si + Si_7$ was never produced. The equilibrium binding energies and associated magic numbers have been shown to be unsatisfactory in describing the behaviour of the clusters under these dynamic conditions. We have introduced the concept of an effective binding energy that incorporates the highly anharmonic dynamics of the clusters. The magic numbers predicted by these effective binding energies are in agreement with both the results of our simulations and experiment. In particular, $Si_4$ and $Si_6$ are both shown to be magic whilst the prediction of the equilibrium values that $Si_7$ is magic is shown to be incorrect. The energy changes predicted using the effective binding energies also account (with the exception of $Si_8$ which is unable to accommodate all of the energy of the collision) for the relative abundances of reaction products observed in the simulations.

ARP acknowledges the financial support of the University of York.

## References

[*] E-mail address: arp28@phy.cam.ac.uk

[†] E-mail address: rwg3@york.ac.uk

**Table I**. Summary of the collisions performed and relative abundances of reaction products.

| Impact parameter, $b$ (Å) | Number of reactive collisions | Reaction products - number of occurrences | | | | |
|---|---|---|---|---|---|---|
| | | $Si_8$ | $Si + Si_7$ | $Si_2 + Si_6$ | $Si_3 + Si_5$ | $Si_4 + Si_4$ |
| 0.00 | 12 | 1 | 0 | 1 | 1 | 9 |
| 0.27 | 12 | 1 | 0 | 0 | 2 | 9 |
| 0.53 | 12 | 6 | 0 | 1 | 1 | 4 |
| 1.06 | 12 | 2 | 0 | 0 | 1 | 9 |
| 1.59 | 12 | 3 | 0 | 0 | 0 | 9 |
| 2.65 | 12 | 3 | 0 | 0 | 1 | 8 |
| 4.76 | 12 | 0 | 0 | 0 | 2 | 10 |
| 5.82 | 9 | 2 | 0 | 0 | 0 | 7 |
| Total No. of reactive collisions: | 93 | | | | | |
| Abundances of reaction products as % of the No. of reactive collisions: | | 19.4% | 0.0% | 2.2% | 8.6% | 69.9% |

**Figure 1.** An illustration of the concept of an effective harmonic potential, used to characterize the dynamic behaviour of a cluster exploring what is actually an anharmonic potential. The potential energy is plotted schematically as a function of atomic configuration.

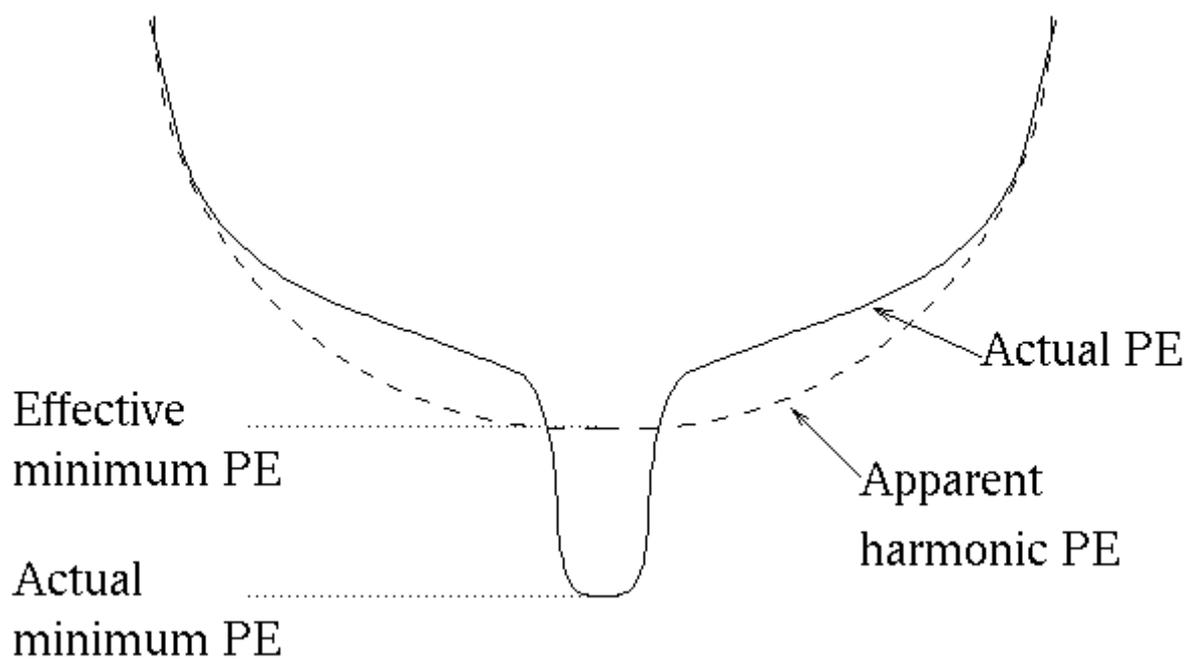

**Figure 2.** The top graph shows the variation of binding energy per atom as a function of cluster size. For the effective binding energy curve, values calculated from thre reaction products were used with the reactant values for Si$_3$ and Si$_5$ shown as additional solid points. An enhanced value compared to neighbouring values indicates that a cluster is more stable than expected and hence 'magic'. The lower plot of the fragmentation energy, BE($N$)−BE($N$−1), emphasizes the magic status of the 4- and 6-atom clusters when the effective binding energies incorporating dynamic effects are used. This is in contrast to the superimposed equilibrium curve which predicts that Si$_7$ is magic.

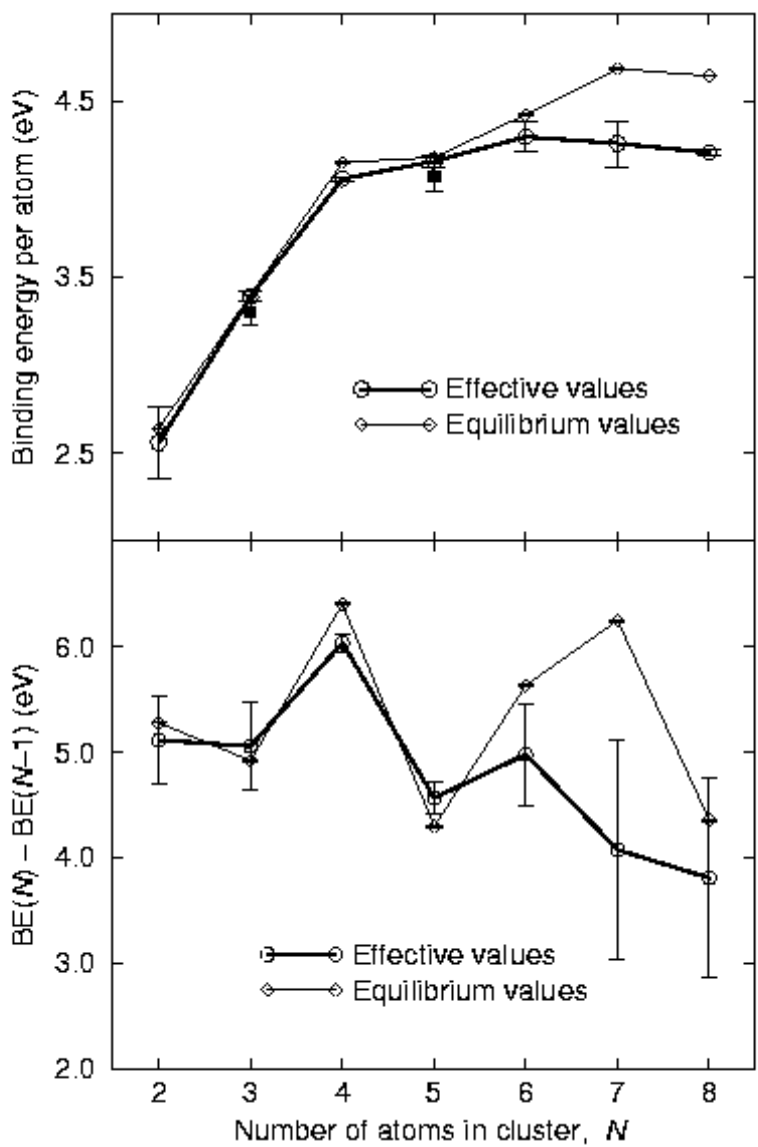

**Figure 3**. The variation of potential energy (top) and mean coordination number (bottom) during an example collision with b=5.82 Å. The collision proper begins at $t$=1.3 ps (first arrow) and is followed by a rapid fall in potential energy as the clusters mix. At $t$=3.8 ps the mean coordination number falls as the $Si_8$ almost splits into $Si_3$ + $Si_5$ before coming back together and finally splitting as two $Si_4$ clusters at $t$=4.8 ps (second arrow). Snapshots of the atomic positions are shown at 0.73, 1.45, 2.42, 3.75, 4.72 and 5.32 ps.

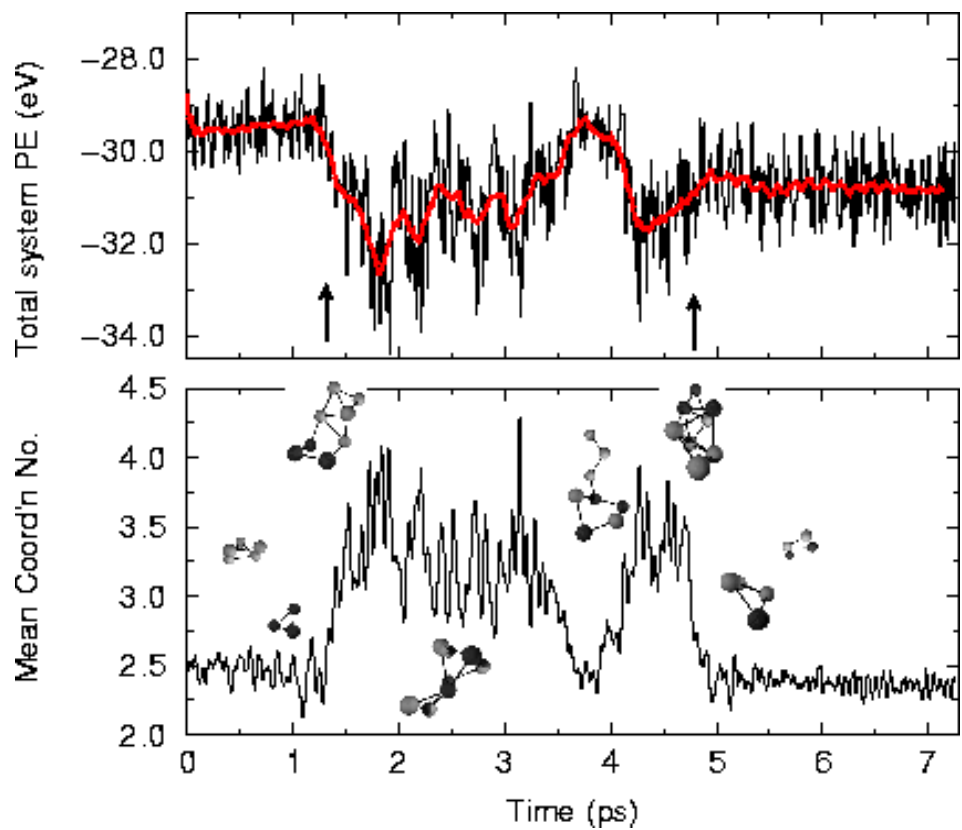